\documentclass{emulateapj}

\usepackage{hyperref}
\usepackage{color,graphicx}
\usepackage{amsmath}

\begin{document}

\title{A detailed statistical analysis of the mass profiles of galaxy clusters}
\author{Ole Host\altaffilmark{1,2}, Steen H.~Hansen\altaffilmark{2}}
\altaffiltext{1}{Department of Physics \& Astronomy, University College London, Gower Street, London, WC1E 6BT, UK}
\altaffiltext{2}{Dark Cosmology Centre, Niels Bohr Institute, University of Copenhagen, Juliane Maries Vej 30, DK-2100 Copenhagen, Denmark}
\submitted{May 5, 2011}

\begin{abstract}
The distribution of mass in the halos of galaxies and galaxy clusters has been probed observationally, theoretically, and in numerical simulations. Yet there is still confusion about which of several suggested parameterized models is the better representation, and whether these models are universal. We use the temperature and density profiles of the intracluster medium as measured by X-ray observations of 11 relaxed galaxy clusters to investigate mass models for the halo using a thorough Bayesian statistical analysis. We make careful comparisons between two- and three-parameter models, including the issue of a universal third parameter. We find that, of the two-parameter models, the NFW is the best representation, but we also find moderate statistical evidence that a generalized three-parameter NFW model with a freely varying inner slope is preferred, despite penalizing against the extra degree of freedom. There is a strong indication that this inner slope needs to be determined for each cluster individually, i.e.~some clusters have central cores and others have steep cusps. The mass-concentration relation of our sample is in reasonable agreement with predictions based on numerical simulations.

\end{abstract}

\maketitle

\section{Introduction}

The potential of gravitationally bound structures in the Universe, ranging in size from dwarf galaxies to galaxy clusters, is sourced by a composite mass distribution of dark matter, baryonic matter in gas form, and collapsed objects such as stars in galaxies and galaxies in clusters. The investigation of these mass distributions entails a number of questions: what is the equilibrium shape of the distributions? Is it universal across ten magnitudes of mass and at all redshifts? Does it depend on cosmology or on the merger history of the individual halos?

The main developments in understanding the distribution of matter in a self-gravitating halo have been found through numerical simulations of the formation of structure in the universe within a given cosmological model. Advances have been achieved through the improvement of numerical codes and the increase of raw computing power on one hand and a more refined understanding of which questions that need to be answered on the other. Perhaps the most fundamental idea that has come out of the numerical approach is that relaxed halos have a number of universal properties, including the distribution of matter \citep{1997ApJ...490..493N,2001ApJ...563..483T,2005ApJ...634..756A} and the dynamical structure \citep{2001ApJ...555..240B,2006NewA...11..333H}. However, the simulations have not been able to reach agreement about the exact behavior of the profiles in the innermost regions, where the limited force resolution sets a lower limit to the radial range that can be probed. In particular, questions have been raised about the value of the logarithmic slope and whether that value is universal or not \citep{1998ApJ...499L...5M,2001ApJ...554..903K,2003ApJ...597L...9Z,2004MNRAS.349.1039N,2004ApJ...606..625F,2006AJ....132.2685M,2006AJ....132.2701G,2006MNRAS.368..518V,2008MNRAS.387..536G}. A further complication arises when the simulations are compared with observations since the gravitational potential of the baryonic component, which is very time consuming to model in the simulations, cannot be neglected in the center. This complication can in principle both change the slope of the dark matter profile as well as alter the total mass profile (\citet{1986ApJ...301...27B,2001ApJ...560..636E,2004ApJ...616...16G,2010MNRAS.408.1998S}. Recently, \citet{2010arXiv1010.2539D} suggested adiabatic contraction as an approach to understanding the actual formation of halos from initially random perturbations.

Theoretically, most efforts have focused on understanding collisionless halos of dark matter only, but this approach is hampered by the fact that, even under the strongest simplifying assumptions, there are not enough constraints to obtain unique solutions to the collisionless Boltzmann equation \citep{1987gady.book.....B} which governs a dark matter structure. Some results have been obtained by considering individual halos from the statistical mechanics point of view (see \citet{2010ApJ...722..851H} and references therein). Another approach is to take phenomenological input from numerical simulations such as the density profile itself, the pseudo-phase space density \citep{2001ApJ...563..483T,2005MNRAS.363.1057D}, or the density slope-velocity anisotropy relation (from which \citet{2006JCAP...05..014H} predict an inner slope of $0.8$), and implement this into a Jeans equation analysis to predict the consequences of the `inspired guess' (see also \citet{2008ApJ...682..835Z} and references therein). Alternatively one can attempt to model the formation history of the halo including major mergers and steady accretion (e.g., \citet{1987ApJ...318...15R,2004MNRAS.352.1109A,2007ApJ...666..181S,2009ApJ...698.2093D}, and references therein). While these approaches typically yield results in rough agreement with simulations, the modeling can also explore the physical connection between the static and dynamic properties of the halo and it can offer physically constrained extrapolations which are not accessible in simulations.

Observationally, there is a strong discrepancy between the numerical results and the inferred mass distributions in dwarf and low surface brightness galaxies, which are much shallower than predicted, the so-called cusp/core problem (see, e.g., \citet{2003A&A...409...53S,2005AJ....129.2119S,2007ApJ...663..948G}). At the opposite end of the mass spectrum, galaxy clusters are typically found to be in rough agreement with the cuspy profiles found in numerical simulations, but with even greater scatter for the inferred inner slope. There is also significant discussion about the type of model and the number of parameters that are necessary in order to obtain an acceptable description of the data. One common method is based on mass modeling through weak or strong gravitational lensing, which can yield results which are in good agreement with numerical simulations \citep{2005ApJ...619L.143B,2006ApJ...642...39C,2008A&A...489...23L,2009A&A...498...37R}, but also profiles that are significantly shallower \citep{2004ApJ...604...88S,2008ApJ...674..711S}. Another method is based on X-ray observations of the intracluster medium (ICM) which is supported against gravitational collapse by its own pressure. Again, authors find a range of inner slopes \citep{2002MNRAS.331..635E,2003ApJ...586..135L,2006ApJ...650..777Z,2007MNRAS.379..209S,2009ApJ...690..154S}. For both lensing and X-ray studies, there is also the question of the stellar mass contained within the massive central galaxy found in most clusters which may affect the cusp/core problem. Most authors also focus on only one or a few clusters, which of course makes it more difficult to assess the universality of the profiles on an observational foundation.

In the present work we take a sample of 11 highly relaxed clusters and use the measurements of the X-ray emitting gas to infer model-independent mass profiles. We then compare with a number of different models that have been applied as mass profiles in the literature, focusing on three key questions: Which parameterized model is the most successful? How many free parameters are needed to describe the data adequately? Is there evidence for a universal inner slope/shape-type parameter? We answer these questions using a detailed statistical analysis based on Bayesian inference where we use the Bayesian evidence (or marginal likelihood) to make judgments about which model is preferred by the data. 

\section{Density profile models}

Most models that are used for the mass distribution in halos have been introduced as fitting formulae applied to the halos found in numerical simulations. Hence these models are not theoretically founded but rather form a basis on which the predictions of numerical simulations can be compared with observations. Almost all models have two free parameters which determine the mass scale and the spatial extent of the halo, and these two parameters are specific to each halo. Some models have one or more additional parameters which determine the shape of the profile, and which may or may not be universal. Here we consider a number of two- and three-parameter models.

\begin{deluxetable*}{lcccc}
\tablewidth{0pt}
\tablecaption{Density profile models\label{tb:models}}
\tablehead{\colhead{Model} & \colhead{$(\alpha,\beta,\gamma)$} &\colhead{$r_{-2}/r_s$} &\colhead{$\rho_{-2}/\rho_0$} & \colhead{$\mu(x=r/r_s)$}}
\startdata
NFW &  $(1,3,1)$ & 1 & $\frac{1}{4}$ & $\ln (1+x)-x/(1+x)$ \\[1mm]

D\&M & $(\frac{7}{9},\frac{31}{9},\frac{4}{9})$ & $\frac{121}{169}$ & $0.0338$ & $\frac{9}{20}(1+x^{4/9})^{-5} $\\[1mm]

Hernquist & $(1,4,1)$ & $ \frac{1}{2} $ & $\frac{16}{27} $ & $x^2/[2(1+x)^2] $\\[1mm]

Moore & $(\frac{3}{2},3,1)$ & $\frac{1}{2}$ & $\frac{8}{3\sqrt{3}}$ & $2\sinh^{-1}(\sqrt{x})-2\sqrt{x/(1+x)} $ \\[1mm] \tableline

slopeNFW & $(\alpha,3,1)$ & $2-\alpha$ & $(2-\alpha)^{-\alpha}(3-\alpha)^{\alpha-3}$ & \nodata\\[1mm]

transNFW & $(1,3,\gamma)$ & $1$ & $\frac{1}{4}$ & \nodata \\[1mm]

S\'ersic & \nodata & 1 & 1 & $8^{-n}e^{2n}n^{1-3n}\gamma(3n,2 n x^{1/n})$

\enddata
\tablecomments{Properties of the density profiles that we consider, including the $(\alpha,\beta,\gamma)$ specification, the relations between $(r_{-2},\rho_{-2})$ and $(r_s,\rho_0)$, and the shape $\mu (r)$ of the mass profile $M(r)=4\pi r_s^3\rho_0\mu (r)$, if analytical. $\gamma(a,x)$ is the lower incomplete gamma function, $\gamma(a,x)=\int_0^x t^{a-1}e^{-t}dt$.}
\end{deluxetable*}

A whole class of models are `double power-laws' which asymptote to power laws at very small and very large radii. These models can conveniently be summarized in Hernquist's $(\alpha,\beta,\gamma)$ parametrization \citep{1990ApJ...356..359H,1996MNRAS.278..488Z},
\begin{equation}
\rho(r)=\rho_0\left(\frac{r}{r_s}\right)^{-\alpha} \left[1+\left(\frac{r}{r_s}\right)^\gamma\right]^{-\frac{\beta-\alpha}{\gamma}},
\end{equation}
where $\rho_0$ and $r_s$ are scaling constants to be determined for each halo individually. The inner power-law slope is $\alpha$ and the outer slope is $\beta$, while the width of the transition region is controlled by $\gamma$. We consider four such two-parameter profiles: the NFW \citep{1997ApJ...490..493N}, the Dehnen-McLaughlin \citep{2005MNRAS.363.1057D}, the Hernquist, and the Moore profile \citep{1998ApJ...499L...5M}. The properties of these models are summarized in Table \ref{tb:models}.

We also consider three three-parameter models: Two are simply generalized NFW profiles where, in the first case, we allow the inner slope $\alpha$ to vary in order to measure the cuspiness of the halos. The second case, transNFW, is also a generalization of the NFW where now the transition parameter $\gamma$ is free. Such a profile can mimic a steeper inner slope by pushing the inner power law behavior closer to the center. The third profile is the S\'ersic (or Einasto) profile \citep{1963BAAA....6...41S,1969Ap......5...67E},
\begin{equation}
\rho(r)=\rho_s\exp\left(-2n\left[\left(\frac{r}{r_s}\right)^{1/n}-1\right]\right),
\end{equation}
where the parameter $n$ determines the shape of the profile. For $n=4$ the de Vaucouleurs' law describing the surface brightness of elliptical galaxies is recovered. The shape parameter is sometimes given as $\alpha_s=n^{-1}$. Recently, the S\'ersic profile has been claimed to provide a better fit than the NFW to Milky Way-sized haloes formed in numerical simulations, and, interestingly, with a shape parameter that varies significantly from halo to halo \citep{2007ApJ...666..181S,2010MNRAS.402...21N}.

We map the scale radius $r_s$ and scale densities $\rho_s$ or $\rho_0$ of each model to the model-independent parameters $r_{-2}$ and $\rho_{-2}$, which are the radius at which the slope of the density profile is $-2$ and the density at that radius, respectively. This mapping makes comparison of the models easier and enables us to use identical priors in the statistical analysis in all models. 

\section{Data analysis}
We revisit the sample of 11 highly relaxed, low redshift ($z<0.1$) galaxy clusters observed with {\it XMM-Newton} which we already used in \citet{2009ApJ...690..358H} to measure the dark matter velocity anisotropy profile for the first time (see also \citet{2007A&A...476L..37H}). The members of this sample were selected to appear close to round on the sky and not have strong features in the temperature and density profiles. The spectral analysis, point-spread-function (PSF) correction, and deprojection of the X-ray data was carried out in \citet{2004A&A...413..415K} and \citet{2005A&A...433..101P}.  The PSF effects degrade the quality of the signal, particularly from the central regions of the clusters, which leads to larger uncertainties on the radial temperature and density profiles. The deprojection method was non-parametric, i.e.~without any parametric modeling of the radial temperature or density profiles. The outcome, and the starting point for the present analysis, was estimates of the ICM temperature $T_i$ and electron number density $n_{e,i}$ with associated uncertainties in six or seven radial bins, for each of the clusters.

Assuming hydrostatic equilibrium, the ICM gas traces the gravitational potential according to \citep{1978A&A....70..677C}
\begin{equation}\label{eq:he}
\frac{k_BT}{\mu m_H}\left(\frac{d\ln n_e}{d\ln r}+\frac{d\ln T}{d\ln r}\right)=-\frac{GM_{\mathrm{tot}}(r)}{r},
\end{equation}
where $\mu=0.6$ is the mean molecular weight of the ICM. We calculate $M_\mathrm{tot}(r_i)$ of each radial bin through a Monte Carlo (MC) analysis in order to propagate uncertainties accurately. In detail, the prescription for each MC realization is as follows: In each bin $i$ the best estimates of $T_i$ and $n_{e,i}$ are added to random numbers drawn from Gaussian distributions representative of the uncertainties $\delta T_i$ and $\delta n_{e,i}$. In order to apply Equation \eqref{eq:he}, we estimate the logarithmic derivative of, e.g., $T$ at the bin-radius $r_i$ by the slope of the unique parabola that passes through $(\ln r_{i-1},\ln T_{i-1})$, $(\ln r_i,\ln T_i)$, and $(\ln r_{i+1},\ln T_{i+1})$. In this way we can calculate the total mass interior to $r_i$ for that data realization. We impose a number of checks to determine if the data realization is physically sensible: the ICM temperature and density must be greater than zero in all bins, the total mass profile must be increasing with radius and the derived total density profile must also be everywhere positive. We also require that the mass contained in dark matter and stars (which is just the total mass with gas mass subtracted) must be positive. If these conditions are not met the entire realization is discarded. This process is repeated until $N=5000$ realizations have been accepted. From these the sample mean of $\ln M_i$ in each bin is determined, as well as the sample covariance matrix with elements
\begin{equation}\label{eq:cov}
C_{ij}=\frac{1}{1-N}\sum_{k=1}^N(\ln M_{ik}-\langle{\ln M_i}\rangle)
(\ln M_{jk}-\langle{\ln M_j}\rangle),
\end{equation}
where $N$ is the number of Monte Carlo realizations. Even though we sample the ICM temperature and density in each bin independently, the covariance matrix is not diagonal since the derivatives and physical consistency checks induce bin-to-bin correlations in the accepted sample. We use the mean and covariance of $\ln M$ rather than $M$ since, by inspection, the former is closer to being Gaussian distributed.

\begin{figure*}[htbp]
\plotone{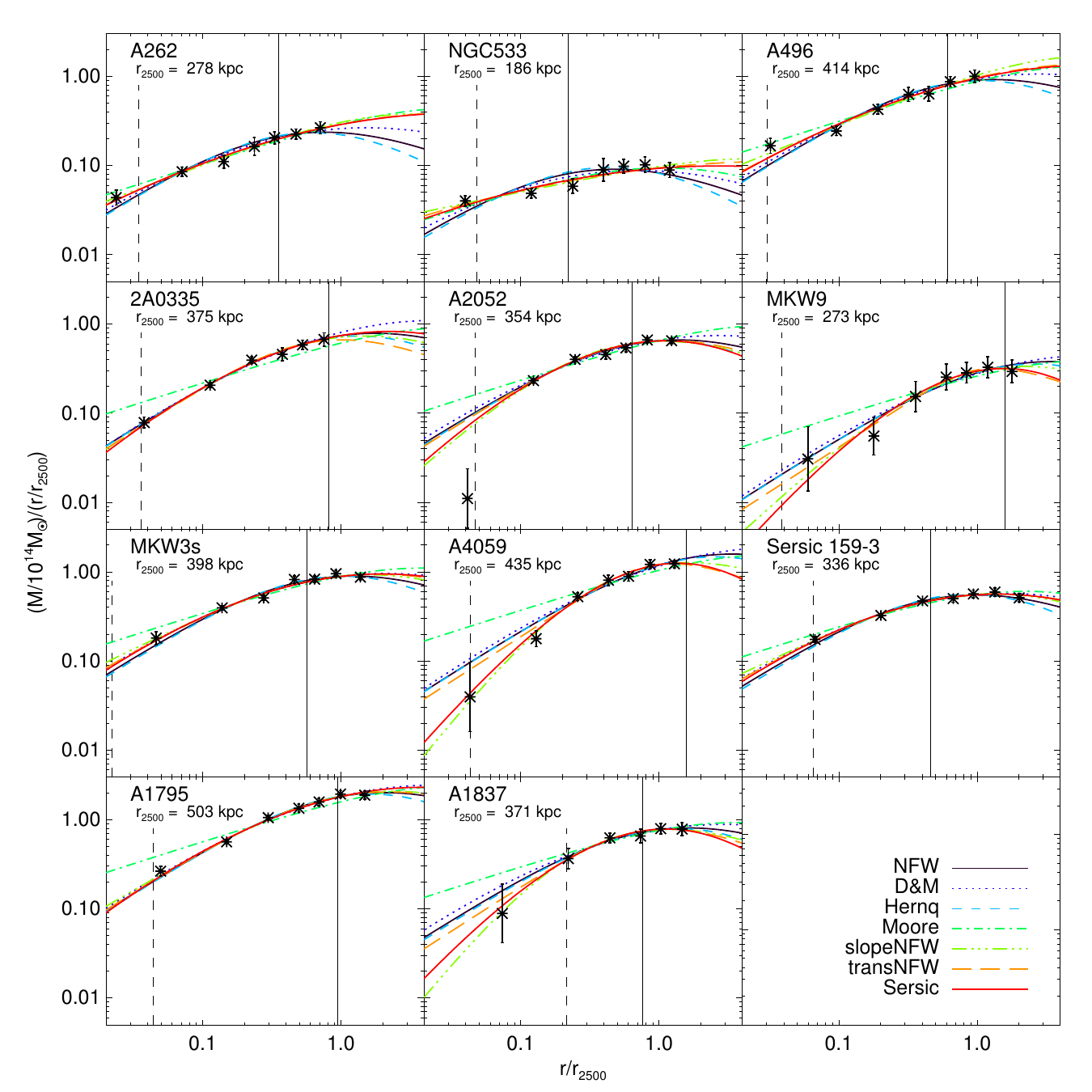}
\caption{Mass profile of each cluster with 68\% uncertainties and best-fit models. The radial axis has been scaled to the best estimate of $r_{2500}$ from the MC analysis, and the mass axis has been scaled by $r^{-1}$.} The dashed vertical lines indicate the estimated extent of the Brightest Cluster Galaxy and the solid vertical lines show the best fit estimate of the NFW scale radius $r_{-2}$. \label{fi:mass}
\end{figure*}

Figure \ref{fi:mass} shows the resulting non-parametric mass profiles measured from the data and their uncertainties, as well as the best fits of the models we consider. Note that the models shown serve only illustrative purposes and the statistical analysis is not limited to the best fits, as discussed below. The vertical dashed line of each panel indicates the $K$-band half-light radius of the central BCG galaxy. The radii are taken from the 2MASS Extended Source Catalog \citep{2006AJ....131.1163S} data products, specifically the de Vaucouleurs' fits to the $K$-band photometry. We take these radii to indicate the likely onset of significant baryonic effects and note that this is typically only an issue in the innermost bin.

\section{Statistical analysis}
We take a Bayesian approach to the statistical analysis and the usual starting point is the likelihood function, which we calculate in the following manner: It requires less manipulation of the data to calculate the mass profile from the observations than to calculate the density profile. Therefore we integrate the density profile analytically or numerically for each model to obtain the mass distribution and compare with the data in mass space, not density space. Further, as mentioned above, we have found in the MC analysis that the mass samplings in each bin are close to being log-normally distributed. Therefore we construct the likelihood $\mathcal{L}(M_i)=\exp(-\chi^2/2)$ from the $\chi^2$ function,
\begin{equation}
\chi^2=\sum_{i,j}(\ln M_i-\ln M(r_i)) C_{ij}^{-1}(\ln M_j-\ln M(r_j)),
\end{equation}
where $M(r_i)$ is the model mass profile at the radial centre $r_i$ of bin $i$, and $\ln M_i$ and $C_{ij}$ are determined by the MC analysis.

The main goal is to decide which model is the better representation of the data. We do this  by calculating the Bayesian evidence of each model, which is a quantitative measure of the agreement between model and data \citep{2008ConPh..49...71T}. First we calculate the likelihoods of each model on a grid in the parameter space $\theta=(\log r_{-2},\log\rho_{-2})$. Next, we construct the posterior probability distribution by combining the likelihood function with a prior probability distribution $\pi(\theta)$ which resembles our knowledge of $\log r_{-2}$ and $\log \rho_{-2}$ before taking the data into account. We discuss the choice of prior below. We then integrate the posterior to obtain the Bayesian evidence,
\begin{equation}
E=\int_\mathrm{all}d\theta \pi(\theta)\mathcal{L}(\theta,\ln M_i),
\end{equation}
which is essentially the weighted average of the likelihood over the prior volume. The evidence of a model, given the data and a prior, quantifies how well that model explains the data. It is important to stress that the comparison is made over all of the parameter space contained by the assumed priors, not just at the best fitting set of parameters. When comparing models the Bayes factor $B_{12}=E_1/E_2$ shows how much more (or less) probable model 1 is than model 2, in light of the data. Traditionally, this is gauged on Jeffrey's scale where a Bayes factor of $\ln B_{12}<1$ is labeled `inconclusive' evidence for model 1 over model 2 while `weak', `moderate', and `strong' evidence corresponds to $\ln B_{12}$ values $<2.5$, $<5$, and $>5$, respectively.

We choose priors which are constant in the logarithms of $r_{-2}$ and $\rho_{-2}$. The flat logarithmic prior is the uninformative prior for scaling parameters \citep{2008ConPh..49...71T} since it reflects ignorance about the magnitude of the parameter. However we restrict the range of the priors, so that we end up with top-hat priors in $\log r_{-2}$ and $\log \rho_{-2}$. As a reference point we first assume a top-hat prior relative to the best estimate of $r_{2500}$ as determined in the MC analysis. The scale radius $r_{2500}$ is defined as the radius within which the mean density is $2500$ times the critical density of the universe. The top-hat prior in $\log r_{-2}$ ranges from $1.5$ magnitudes below $r_{2500}$ to 0.5 above. The basic idea behind this prior is that the transition or `roll' of a model should occur close to $r_{2500}$, as it does in haloes in numerical simulations, and also to prevent the model from behaving as a simple power-law by pushing the transition from the inner to the outer power law far away from the range of the data. We emphasize that this is still a conservative prior, as current simulations typically resolve 2--3 radial orders of magnitude with $r_{-2}$ located about one order of magnitude below the virial radius \citep{2001MNRAS.321..559B}. The prior in $\log \rho_{-2}$ is also a top-hat in the range $10^{-26}$--$10^{-21}\,$kg$\,$m$^{-3}$, which in practice means that the likelihood is vanishingly small at the boundaries of the prior. 

\begin{deluxetable}{lrrrr}
\tablewidth{0pt}
\tablecaption{Bayes Factor $\ln B$ for the two-parameter models, relative to the NFW profile\label{tb:bayes2}}
\tablehead{\colhead{Cluster} & \colhead{$z$} &\colhead{D\&M} &\colhead{Hernq.}&\colhead{Moore}}
\startdata
A262 		& 0.015 & -2.1 & 0.9 & -2.9 \\
NGC533 		& 0.018 & -1.8 & 1.2 & -3.1 \\
A496 		& 0.032 & -1.4 & 0.6 & -0.8 \\
2A0335+096 	& 0.034 & 0.5 & -0.1 & 14.7 \\
A2052 		& 0.036 & 1.9 & -0.3 & 6.6\\
MKW9 		& 0.040 & 0.5 & -0.1 & 1.5 \\
MKW3s 		& 0.046 & 1.8 & -0.2 & 6.7 \\
A4059 		& 0.047 & 1.6 & -0.4 & 10.8 \\
S\'ersic 159--3 	& 0.057 & -0.5 & 1.7 & 3.4 \\
A1795 		& 0.064 & 2.6 & -0.6 & 20.8 \\ 
A1837 		& 0.071 & 0.5 & -0.2 & 1.4 \\ 
\tableline
Total & \nodata & 3.8 & 2.6 & 59\phd\phn
\enddata
\tablecomments{A positive value of $\ln B$ indicates that the NFW profile is preferred over the considered model. Note that this does not imply any bias towards the NFW as the Bayes factor of any two other models is just the difference between the respective Bayes factors given here.}
\end{deluxetable}

\subsection{Two-parameter model results}

The result of the model comparison is summarized in Table \ref{tb:bayes2}, where the NFW model is compared against each of the other two-parameter models. A positive Bayes factor indicates that the NFW model is preferred. This does not imply any bias on the NFW since any two models can be compared by subtracting the Bayes factors we give for them from one another. We find that, individually, the clusters yield strong constraints only against the Moore model, while the evidences for or against the D\&M and Hernquist models are either weak or inconclusive on Jeffrey's scale. If instead we consider the cumulative Bayes factor summed over the full sample, the NFW is found to be the preferred model overall, i.e., as a universal two-parameter profile our sample favors the NFW model. The Hernquist profile and the D\&M profile are weakly and moderately disfavored, respectively, with cumulative Bayes factors of 2.6 and 3.8 while the Moore profile is convincingly ruled out with a factor of 59. The weak constraint on the Hernquist profile is not surprising as data extending out to the virial radius would likely be needed to properly distinguish this model from the NFW. 

In Table \ref{tb:priors} we present the effects of varying the priors. The evidence against the D\&M profile becomes strong when we limit the range of the prior in $\log r_{-2}$ to the smaller interval $(-0.75,0.25)$. The same is true if we choose top-hat priors in $(r_{-2},\rho_{-2})$ instead of the logarithmic priors. Finally, the D\&M model is also disfavored slightly more if we apply a `soft' Gaussian prior in $\log r_{-2}$. The Bayes factor for the Hernquist model is robust under the same variations, while the Moore profile is very strongly ruled out in all cases. We conclude that our two-parameter model selection results are stable against variation amongst reasonable choices of priors, which means that the data are of sufficient quality to make robust conclusions.

A more interesting issue to consider than the priors is that the preference for the NFW profile over the Hernquist and D\&M profiles is somewhat susceptible to `jackknife' resampling: if we recompute the cumulative Bayes factor eleven times systematically leaving a single cluster out each time, then there are a few cases where the strength of the evidence is reduced to inconclusive but also cases where it is increased to strong (against the D\&M). This is largely due to the fact that our data sample is somewhat inhomogeneous in terms of the relative statistical uncertainty on the mass profile. For example, a comparison of the error bars of MKW9 with those of A1795 or S\'ersic 159-3 (see Figure \ref{fi:mass}) immediately shows that the former is much less constraining than the latter two. This means that our sample is a mixture of strongly and weakly constraining clusters and this is reflected in Figure \ref{fi:chart} where the contributions from individual clusters clearly varies. There appears to be a trend that the clusters A262, NGC533, and A496, which are the lowest redshift and some of the least massive in our sample, stand out by preferring the D\&M and the Moore profile. However, such trends may just as likely be spurious effects caused by the relatively small sample size. The D\&M profile can easily be preferred by clusters that also prefer the Moore profile since, by extending the transition region, the D\&M profile can push the inner asymptotic power law well inside the radial range of the data.

Finally we note that the minimum $\chi^2$ values for the models support the more detailed analysis: for 53 degrees of freedom we get minimum $\chi^2$'s of 84 for the NFW, 97 for the D\&M, 86 for the Hernquist, and 208 for the Moore profile. Major contributions to these $\chi^2$ values come from the two clusters MKW3s with $\chi^2=13.5$ and A4059 with $\chi^2=13.7$ for the NFW model and similar or larger values for the other models. The corresponding $p$-values imply that the D\&M $\chi^2$ is about 20 times less likely to have occurred by chance (if the D\&M model is correct) than the NFW model is (if the NFW model is correct). Compare this with the Bayesian odds that the NFW is $\sim40$ times more probable than the D\&M. Note that the actual best-fits are slightly smaller since we evaluate the $\chi^2$ on a grid instead of minimizing it with a dedicated search. The $\chi^2$ values also show that, in terms of goodness-of-fit, our sample is rather inhomogeneous. 

\begin{deluxetable}{lrrrr}
\tablewidth{0pt}
\tablecaption{Total Bayes factor $\ln B$ for the two-parameter models assuming various priors, relative to the NFW profile\label{tb:priors}}
\tablehead{\colhead{Prior} &\colhead{Range $\log r_{-2}$}&\colhead{D\&M} &\colhead{Hernq.}&\colhead{Moore}}
\startdata
Top-hat in $\log r_{-2}$	& (-1.5,0.5) & 3.8 & 2.6 & 59\\
Top-hat in $\log r_{-2}$	& (-3,3)	  & 3.0 & 2.6 & 42\\
Top-hat in $\log r_{-2}$	& (-0.75,0.25)	  & 7.8 & 2.4 & 74\\ 
Top-hat in $(r_{-2},\rho_{-2})$	& (-1.5,0.5) & 9.2 & 1.4 & 68\\ 
Gaussian in $\log r_{-2}$ & \nodata & 5.3 & 2.4 & 57 
\enddata
\tablecomments{The top line is the fiducial prior used in Table \ref{tb:bayes2}. In the next two cases the range of the prior in $\log r_{-2}$ (in units of $r_{2500}$, see text) is varied, and in the following case top-hat priors in  both $r_{-2}$ and $\rho_{-2}$ are applied. The final case assumes a Gaussian prior in $\log r_{-2}$ with mean -0.25 and width 0.5.}
\end{deluxetable}

\subsection{Three-parameter model results}

\begin{figure*}[htbp]
\begin{center}
\plotone{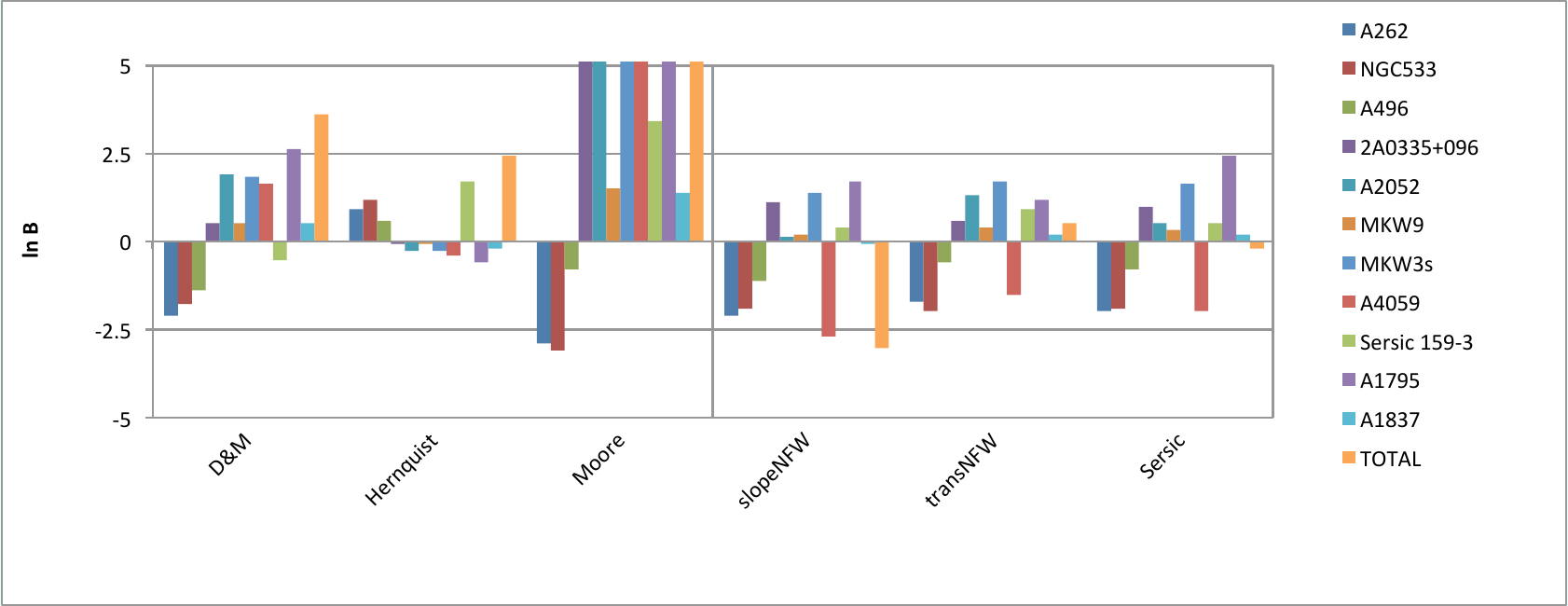}
\caption{Bar chart of the Bayes factors $\ln B$ for the various models considered, relative to the NFW, as given in Table \ref{tb:bayes2} and \ref{tb:bayes3}. The Bayes factors are additive so the contribution of individual clusters to the total Bayes factor is easily assessed. The values shown are based on the fiducial priors discussed in the text.}\label{fi:chart}
\end{center}
\end{figure*}

For the three-parameter models we again want to test whether the models represent the data better than the NFW. In this case the comparison is slightly more involved to evaluate since there is the freedom of an additional parameter to take into account. This naturally yields a lower value of the evidence if the extra parameter does not provide a better description of the data, or, to put it otherwise, the third parameter must improve the fit over a significant volume of parameter space in order to be preferred over the NFW.
It is important to stress that there is no assumption about the third parameter being universal. On the contrary, we ask whether the data require the additional freedom of an extra parameter which is determined individually for each cluster.

The model comparison proceeds as before: we calculate the evidence for each of the three-parameter models with the same priors in $\log r_{-2}$ and $\log \rho_{-2}$ as in the fiducial two-parameter analysis for all models. For the slopeNFW we choose a top-hat prior for $\alpha$ which ranges from 0 to 1.75, i.e.~from a cored profile to a profile slightly steeper than the Moore profile. We do not want to go all the way to $-2$ since $r_{-2}$ tends to zero and eventually becomes undefined as $\alpha$ approaches $-2$. For the transNFW, we choose a logarithmic prior with $\gamma$ in the range $(0.1,4)$ which allows this profile to mimic a steeper inner profile by pushing the asymptotic inner power law inside the radial range of the data. Finally, we take a logarithmic prior for $n$ in the range (2,15) for the S\'ersic model, motivated by numerical simulations which have best fits S\'ersic profiles with $n$=5--9. The logarithmic prior also has the advantage that it is invariant whether one prefers $n$ or $\alpha_s=1/n$ as the parameterization.

The resulting Bayes factors relative to the NFW are given in Table \ref{tb:bayes3} and summarized in the chart in Figure \ref{fi:chart}. The individual clusters provide only weak evidence for or against any of the models. Based on the whole sample, the model selection is inconclusive for the transNFW and S\'ersic models but there is `moderate' evidence for the slopeNFW model over the NFW with a Bayes factor of $-3.1$. This corresponds to odds of 22 to 1 in favor of the slopeNFW model and shows that, overall, the slopeNFW has the highest evidence of all models considered. The data show a moderate need for a free inner slope, despite the penalty against the extra freedom built into the Bayesian analysis. Most of the discriminatory power is carried by a few clusters such as NGC533, A4059, and A1795 and removal of any of these clusters from the sample would change the Bayes factor significantly. Therefore we caution that the moderate preference for the slopeNFW model is somewhat susceptible to selection effects since the constraints from individual clusters vary in strength. We also find some sensitivity to the choice of prior: if the upper bound of $\alpha$ is extended from 1.75 up to 1.9, the Bayes factor for the slopeNFW model changes to $-2.3$, while if it is set to the Moore profile at $1.5$ the Bayes factor becomes $-3.6$. If the lower bound of $\alpha$ is increased to $0.5$, the Bayes factor remains unchanged at $-3.1$. These results are largely caused by a few clusters which prefer steep inner slopes, as will be shown below.

\begin{deluxetable}{lrrr}
\tablewidth{0pt}
\tablecaption{Bayes Factor $\ln B$ for the three-parameter models, relative to the NFW profile.\label{tb:bayes3}}
\tablehead{\colhead{Cluster} & \colhead{slopeNFW}&\colhead{transNFW}&\colhead{S\'ersic}}
\startdata
A262 		& -2.1 & -1.7 & -2.0 \\
NGC533 		& -1.9 & -2.0 & -1.9 \\
A496 		& -1.1 & -0.6 & -0.8 \\
2A0335+096 	&  1.1 & 0.6 & 1.0 \\
A2052 		&  0.1 & 1.3 & 0.5 \\
MKW9 		& 0.2 & 0.4 & 0.3 \\
MKW3s 		&  1.4 & 1.7 & 1.6 \\
A4059 		&  -2.7 & -1.5 & -2.0 \\
S\'ersic 159--3 	&  0.4 & 0.9 & 0.5 \\
A1795 		&  1.7 & 1.2 & 2.4 \\ 
A1837 		& -0.1 & 0.2 & 0.2 \\ \tableline
Total 		& -3.1 & 0.6 & -0.2
\enddata
\tablecomments{A positive value of $\ln B$ indicates that the NFW profile is preferred over the considered model. A top-hat prior in $\log r_{-2}$ of $(-1.5,0.5)$ around the best estimate of $r_{2500}$ for each cluster is assumed.}
\end{deluxetable}

\subsection{Constraints on the third parameters}

\begin{figure*}[htbp]
\begin{center}
\includegraphics[width=.32\textwidth]{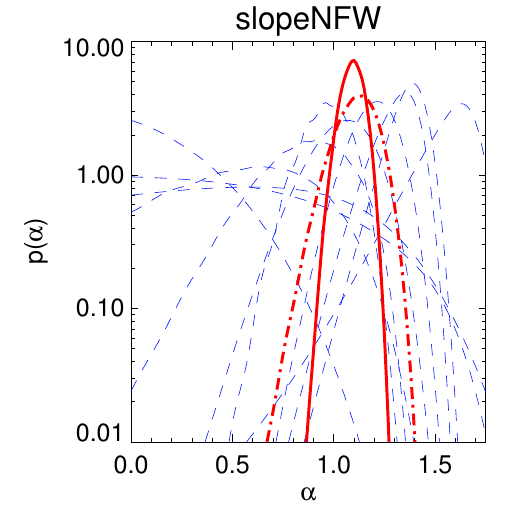}
\includegraphics[width=.32\textwidth]{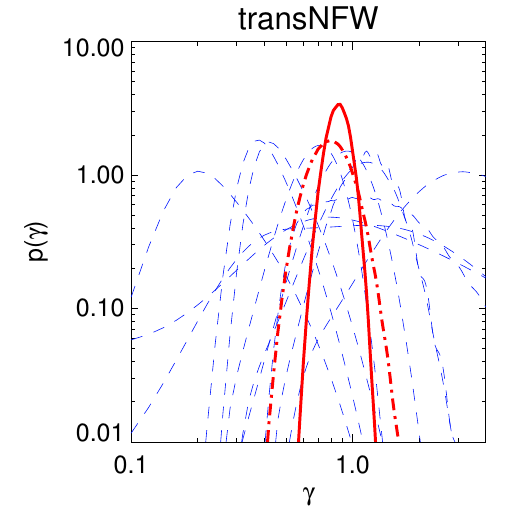}
\includegraphics[width=.32\textwidth]{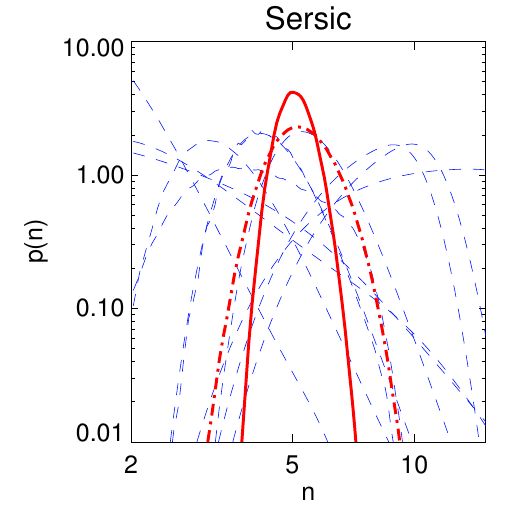}

\caption{Probability distributions for the third parameter in each of the three-parameter models: slopeNFW $\alpha$ (left), transNFW $\gamma$ (middle), and S\'ersic $n$ (right). In each panel, the full line shows the joint posterior for all clusters combined while the dot--dashed line shows the joint posterior obtained using the method of hyper-parameters (see text). The dashed lines show the pdf's of individual clusters. Note that each posterior is normalized to unity so it is not possible to draw conclusions about the quality of fit of the individual clusters from this plot. The standard 95\% credible intervals are $(0.98,1.19)$ for $\alpha$, $(0.70,1.08)$ for $\gamma$, and $(4.3,6.1)$ for $n$. With the hyper-parameters, the intervals are instead $(0.91,1.30)$ for $\alpha$, $(0.52,1.20)$ for $\gamma$, and $(3.8,7.4)$ for $n$. We assume top-hat priors in $\alpha$, $\ln \gamma$, and $\ln n$.}\label{fi:third}
\end{center}
\end{figure*}

Finally, for the three-parameter models we also want to constrain the preferred value of the third parameter. Unlike above, this analysis assumes that there is a universal value for the third parameter and attempts to identify that value. We use the same priors as in the previous analysis for the third parameter of each model, but now we marginalize over  $(\log r_{-2}, \log \rho_{-2})$ to find the one-dimensional posterior probability distribution for the third parameter for each cluster. Then we combine the results from the individual clusters into a joint posterior which is simply the product of the the individual ones. We calculate 95\% credible intervals for both the individual and the joint posterior. However, we know from the previous analysis that each three-parameter model is preferred by some clusters but not by others. Therefore we also use the method of hyper-parameters \citep{2000MNRAS.315L..45L} which allows the contribution from individual data-sets to the joint posterior to be weighted. These weights are marginalized over assuming logarithmic priors with the result that in the joint likelihood one replaces
\begin{equation}
\sum_i\chi_i^2\rightarrow \sum_i N_i\ln\chi^2_i,
\end{equation}
where $N_i$ is the number of data points in data-set $i$. The result is that clusters that are not described well by the model do not constrain the parameters as strongly as clusters that are well described. The price to pay is that the effective sample size is reduced which, all else being equal, will lead to wider and more conservative credible intervals.

The results are shown in Figure \ref{fi:third}, where in each panel the fully drawn line is the joint posterior, the dotted line is the hyper-parameters posterior, and the dashed lines are the posteriors of the individual clusters. The generalized NFW models are slightly different from, but not in disagreement with, the NFW with 95\% credible intervals of $(0.98,1.30)$ for $\alpha$ and $(0.70,1.08)$ for $\gamma$. The interval for the S\'ersic $n$ parameter is $(4.3,6.1)$, in good agreement with the values reported by the Aquarius numerical simulations for Milky Way-sized halos \citep{2010MNRAS.402...21N}. The intervals derived using the method of hyper-parameters are wider, as expected: $(0.91,1.30)$ for $\alpha$, $(0.52,1.20)$ for $\gamma$, and $(3.8,7.4)$ for $n$. The difference between the hyper-parameters method and the conventional calculation illustrates the need for a cautious approach to in-homogeneous data-sets. We believe the hyper-parameters method yields the more trustworthy results in the case at hand, while on the other hand we acknowledge that they are not very constraining.

An inspection of the contribution from individual clusters reveals some issues: It is clear that for each model a number of clusters provide very little information about the third parameter, i.e.~the model describes the mass profile almost equally well regardless of the third parameter value. This is actually expected, given the varying size of the Bayes factors in table \ref{tb:bayes3}. There are also a few cases, particularly for the transNFW model, where the posterior peaks very close to or on the bounds of the prior. In such cases the results, e.g.~the individual credible intervals, are of course very prior-dependent which again indicates that the data are not very discriminatory with respect to the prior. On the other hand, rather drastic priors or small sub-samples must be used in order to significantly affect the credible intervals of the joint posterior, especially for the hyper-parameters method.

\begin{figure}[bt]
\includegraphics[width=\columnwidth]{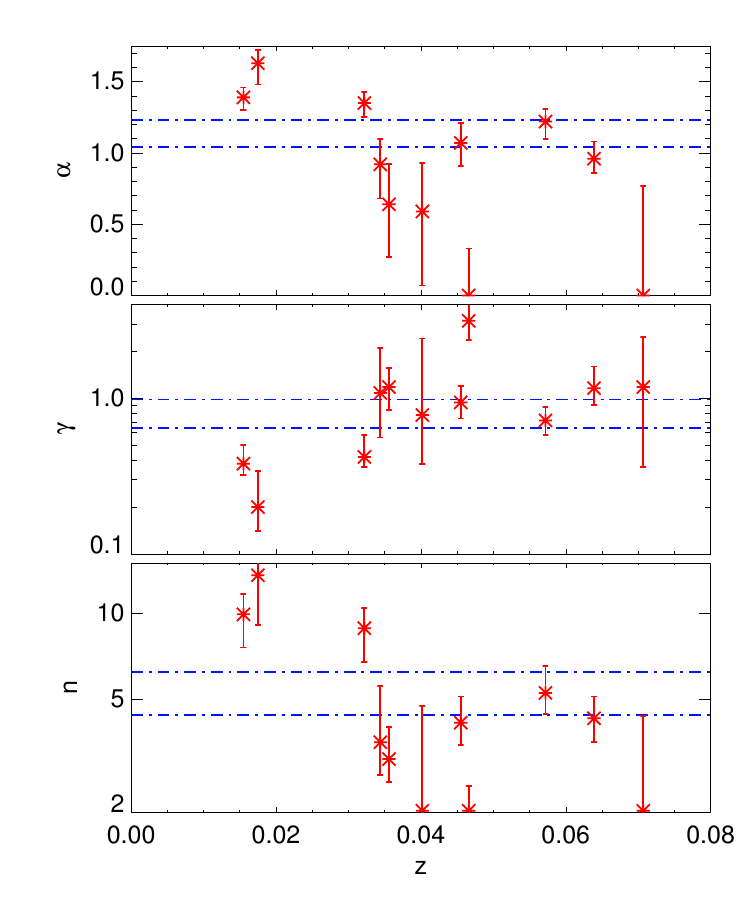}\label{fi:zdep}
\caption{The individual clusters' constraints on the third parameter in each of the three-parameter models. In this case we show the 68\% credible intervals, and the horizontal lines indicate the 68\% range of the joint posterior calculated using the method of hyper-parameters. Refer to Table \ref{tb:bayes2} for the redshifts of each cluster.}
\end{figure}

Figure \ref{fi:zdep} shows the individual clusters' constraints on $\alpha$, $\gamma$, and $n$. There is perhaps the slightest of hints of a redshift--dependence in the constraints but the sample size does not allow us to probe such an issue in detail. It should also be noted that any hint of a redshift--dependence could actually be caused by a mass--dependence instead since, e.g., the two clusters at the lowest redshifts are also the least massive. 

A different picture emerges when we consider the overlap of the individual clusters' credible intervals for the slopeNFW model. For example, no value of $\alpha$ is contained in all 11 95\% credible intervals, and only the very short range $(1.08,1.10)$ is contained in all but two intervals. Likewise the NFW $\alpha=1$ case is excluded from four of the eleven intervals. These results, as well as a visual inspection of Figure \ref{fi:third}, do not seem to support a universal shape parameter. The situation is not quite as compelling for the transNFW and S\'ersic models which is likely the reason that they do not stand out from the NFW in the model selection. In fact, we believe it is a reasonable statement that the success of the slopeNFW model is precisely due to the very different preferred values of $\alpha$ from cluster to cluster. This puts a strong question mark against the idea of universal third parameter.

We conclude that there is moderate evidence for the slopeNFW model to be preferred over the simple NFW, while the transNFW and S\'ersic models do not stand out against the two-parameter NFW profile. If the inner slope of the slopeNFW model is universal, we constraint it to be close to $-1$ but preferably slightly steeper. However, the spread of the individual clusters' preferred ranges suggests that the inner slope is not universal.

\section{Biases}
So far we have discussed the interpretation of our results with respect to the statistical evidence. However, a number of biases, or systematic uncertainties, can be thought of that may affect our results. Loosely, these can be grouped into biases that affect both the individual cluster mass modeling and the combined analysis, and selection effects that only influence the latter.

The analysis rests on the ability to produce deprojected temperature and density profiles with uncertainties that correctly mirror the uncertainties in the spectral analysis of the X-ray data. This has been discussed extensively in \citet{2004A&A...413..415K}.
The basic assumption in determining the mass distribution of a galaxy cluster is that the cluster is relaxed, and obeys the equation of hydrostatic equilibrium. Numerical simulations indicate that the additional pressure associated with turbulence and bulk motion in the ICM yields an underestimate of the mass in the region of $5-20$\% with the larger values corresponding to large radii, $r_{500}$ and greater \citep{2007ApJ...655...98N,2008A&A...491...71P,2009ApJ...705.1129L}. We do not expect this bias to exceed 10\% in the present case since we do not model further out than to $\sim r_{2500}$. On the other hand, the same numerical simulations indicate that if the turbulent pressure is accounted for, an accurate mass reconstruction is possible. This point demonstrates that deviations from spherical symmetry are not a major concern in the error budget. 

A related question is whether the parameterized profiles should be tested against the total mass distribution or the dark matter mass profile only. While the predictions of numerical simulations are founded in dark matter-only simulations, it is not clear how much a simulated dark matter-only mass profile is modified by the presence of baryons. Observationally, the ICM contributes about $5-15$\% of the total density in a cluster, again increasing with radius in the range of interest here, so formally there is a difference between the total and the dark matter profile's radial dependence. To test the impact of this, we have rerun the statistical analyses described above, but with an additional step in which the ICM mass is subtracted from the mass estimate of Equation \eqref{eq:he} so that we compare the parametric models to the mass profile of dark matter and stars. We find only minor differences: For the two-parameter models, the total Bayes factors relative to the NFW profile assuming the fiducial prior as in Table \ref{tb:bayes2} are 3.6 (D\&M), 2.4 (Hernquist), and 51 (Moore), i.e.~there is no significant change in the interpretation of the results. For the three-parameter models, the total Bayes factors become $-2.6$ (slopeNFW), $0.7$ (transNFW), and $-0.1$ (S\'ersic), which are in good agreement with the results in Table \ref{tb:bayes3}. Finally, the constraints on the third parameters for the three-parameter models are effectively unchanged. The fact that the results are virtually identical is not surprising since the ICM is a minor and smoothly distributed contribution to the mass profile.

We have also attempted to account for the stellar mass. This is a very subdominant component except close to the center of a cluster where the mass of the BCG galaxy can be significant. The likely range where the BCG should be accounted for is indicated in Figure \ref{fi:mass} and can clearly affect the measurement of the inner slope of the dark matter mass profile. We have attempted to account for the BCG mass by first using the 2MASS $K$-band photometry to determine the luminosity profile \citep{2001ApJ...560..566K} and then convert this to a mass profile by assuming a mass-to-light ratio. \citet{2009MNRAS.394..774L} find that the $K$-band mass-to-light ratio is about unity. This prescription yields a rough and very model-dependent estimate of the stellar mass but it is useful to investigate the feasibility of measuring the dark matter mass distribution. We immediately find consistency issues, however, since it turns out that the total mass profiles we have measured for A2052 and A4059 cannot accomodate such a stellar mass component unless we reduce the $M/L$ ratio significantly. For A2052, we suspect that the mass measured in the innermost radial bin is a low outlier since all models predict a significantly greater mass in that bin, but that does not help us in determining a physically consistent mass model with both stars, ICM gas and dark matter. We conclude that  a more detailed measurement of the BCG stellar mass is necessary to separate that from the dark matter. Such a measurement is outside the scope of this work, so we have restricted ourselves to determining the total mass profiles.

The fact that our results are stable whether we test the mass models against the total or ICM-subtracted mass profiles allows us to gauge how important the mass bias caused by turbulent pressure is. The point is that the turbulent pressure is expected to contribute the same amount (or less) to the total mass estimate as the ICM mass: both contributions are at the $5-15$\% level and radially increasing, and at the maximum radius we consider $\sim r_{2500}$ the gas fraction ($\sim~10\%$) is likely larger than the pressure bias. Since our results are the same whether we account for ICM mass or not, we conclude that this systematic offset is likely much smaller than the statistical uncertainty.

\section{Mass--concentration relation}

\begin{figure}[bt]
\includegraphics[width=\columnwidth]{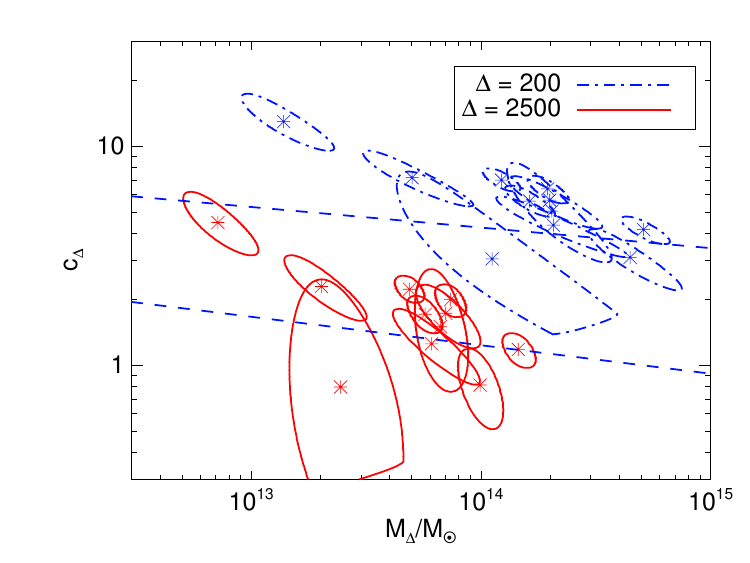}
\caption{The mass--concentration relation of our sample, calculated within the NFW model. The contours contain 95\% of the posterior PDF and are based on the fiducial prior. We show two contours for each cluster: $(M_{2500},c_{2500})$ (red, full lines) which are derived within the radial range of the data, and $(M_{200},c_{200})$ (blue, dot-dashed) which is based on an NFW model-dependent extrapolation to $r_{200}$. The dashed lines show the mean relations for the two values of $\Delta$ from the N-body simulations of \citet{2008MNRAS.391.1940M}, based on the WMAP5 cosmology. The relations are $\log c_{200}=0.83-0.094\log(M_{200}/10^{12}M_\sun)$ and $\log c_{2500}=0.35-0.130\log(M_{2500}/10^{12}M_\sun)$. Given the low redshift of our sample, we have not made any correction for a redshift evolution of $c_\Delta$.}\label{fi:mc}
\end{figure}

An important consequence of the `bottom-up' scenario of structure formation is that smaller halos are denser in the center, since they formed earlier when the density of the Universe was higher. This effect is observed in numerical simulation and it can be expressed as a relation between the halo mass and the concentration parameter. The concentration parameter is defined for a given overdensity as $c_\Delta=r_\Delta/r_{-2}$ (often $r_s$ is used instead of $r_{-2}$ but for the NFW this is unimportant). Simulations usually consider the mass--concentration relation at the virial radius $r_{200}$ but as discussed above we can only reach that radius by model-dependent extrapolation. Therefore, in Figure \ref{fi:mc}, we show the mass--concentration relation of our sample calculated within the NFW model at both $r_{2500}$ and extrapolated to $r_{200}$. Note that our sample is not necessarily representative of the population of clusters.

As can be seen in Figure \ref{fi:mc}, our sample is not ideally suited to derive a relation from, given that six sample members cluster at almost identical values of $M_\Delta$. Instead we compare with the mass--concentration relation of the dark matter-only simulations presented in \citet{2008MNRAS.391.1940M}, which are in reasonable agreement with our sample. We emphasize that the orientation of the uncertainty ellipses is related only to the parameter degeneracies present in the combination of model and mass profile data and has nothing to do with the slope of the mass--concentration relation. The agreement between our observed mass--concentration relation and the predictions of numerical simulations resembles the recent X-ray analysis \citet{2007ApJ...664..123B}, but stands out from the significant discrepancy of the lensing study \citep{2008ApJ...685L...9B}.

\section{Summary \& discussion}
We have conducted a careful statistical analysis of the constraints on mass distribution models of galaxy clusters which can be derived from X-ray observations. We find that the NFW model is the preferred two-parameter model and that the Moore model is decisively ruled out. There is moderate evidence that the data require an additional free parameter that alters the shape of the mass profile and according to our analysis the best choice is a model similar to the NFW but with a freely varying inner slope. If we assume this slope to be universal, we can constrain it to be close to that of the NFW but our data suggest that the shape parameter must be determined individually. 

Significantly, the clusters in our sample prefer quite different values for the inner slope, some prefer flat cores while others prefer steep cusps. The shape-parameters of the S\'ersic and transNFW models also show considerable scatter across our sample. We conclude that there is a strong indication in our data that the total mass profile is not universal but suffers considerable halo-to-halo scatter. The limited size of our sample means that we cannot state whether this is in disagreement with the results of numerical simulations. However, when the goodness--of--fit of each cluster is taken into account using the method of Bayesian hyper-parameters, the credible interval becomes significantly larger, partly due to the smaller effective sample size, but also because of the lack of universality. Alternatively if, against best efforts, some clusters in our sample are not relaxed, that may cause the lack of universality we find. 

This analysis stands out from the numerous observational results that claim significant discrepancies from simulations based on only one or a few observed clusters. We acknowledge that our sample size is still limited, but it allows us to discuss the issue of universality. Given that halos in numerical simulations which include baryons are still not readily mass produced with sufficient resolution, which makes the question of halo to halo scatter difficult to assess, it is not possible to decide if the indication of a non-universal model that we see is at odds with the numerical predictions, nor to assess how the central galaxy affects the simulated mass profiles.

\acknowledgments{We thank Rocco Piffaretti for sharing data with us and for comments on the manuscript, and Andrea Macci\`o  for providing the mass-concentration relation of his numerical simulations. The Dark Cosmology Centre is funded by the Danish National Research Foundation. This publication makes use of data products from the Two Micron All Sky Survey, which is a joint project of the University of Massachusetts and the Infrared Processing and Analysis Center/California Institute of Technology, funded by the National Aeronautics and Space Administration and the National Science Foundation. This research has made use of the NASA/IPAC Extragalactic Database (NED) which is operated by the Jet Propulsion Laboratory, California Institute of Technology, under contract with the National Aeronautics and Space Administration. }

{\it Facilities:} \facility{XMM}

\end{document}